\documentclass[twocolumn]{aastex62}
\usepackage{newtxtext,newtxmath}
\usepackage[utf8]{inputenc} 
\usepackage{color} 
\usepackage[T1]{fontenc}
\usepackage{ae,aecompl}
\usepackage{graphicx}	
\usepackage{amsmath}	
\usepackage{amssymb}
\usepackage{comment}

\graphicspath{{./}{figures/}}

\submitjournal{ApJ}
\shorttitle{DM velocity dispersion and the Cosmic Dawn}

\begin{document}

\title[DM velocity dispersion]%
{Constraints on the velocity dispersion of Dark Matter
from Cosmology and new bounds on scattering from the Cosmic Dawn.}

\correspondingauthor{Iván Rodríguez-Montoya}
\email{irodriguez@inaoep.mx}

\author{Iván Rodríguez-Montoya}
\affiliation{Consejo Nacional de Ciencia y Tecnología.
			Av. Insurgentes Sur 1582, 03940, Ciudad de México, México}
\affiliation{Instituto Nacional de Astrofísica, Óptica y Electrónica.
			Apdo. Post. 51 y 216, 72000. Puebla Pue., México }

\author{Vladimir Ávila-Reese}
\affiliation{Instituto de Astronomía, Universidad Nacional Autónoma de México, Apdo. Post. 70-264, 04510 Ciudad de México, México}

\author{Abdel Pérez-Lorenzana}
\affiliation{Departamento de Física, Centro de Investigación y de Estudios Avanzados del I.P.N. 
       Apdo. Post. 14-740, 07000, Ciudad de México, México.}

\author{Jorge Venzor}
\affiliation{Departamento de Física, Centro de Investigación y de Estudios Avanzados del I.P.N. 
       Apdo. Post. 14-740, 07000, Ciudad de México, México.}

\begin{abstract}
The observational value of the velocity dispersion,
$\Delta\upsilon$, is missing in the Dark Matter (DM) puzzle.
Non-zero or non-thermal DM velocities can drastically
influence Large Scale Structure and the 21-cm temperature
at the epoch of the Cosmic Dawn,
as well as the estimation of DM physical parameters,
such as the mass and the interaction couplings.
To study the phenomenology of $\Delta\upsilon$
we model the evolution of DM in terms of a simplistic and generic 
Boltzmann-like momentum distribution.
Using cosmological data from the Cosmic Microwave Background,
Baryonic Acoustic Oscillations, and Red Luminous Galaxies,
we constrain the DM velocity dispersion for a broad range of masses
$10^{-3} \text{ eV} < m_\chi < 10^9 \text{ eV}$, finding
$\Delta\upsilon_0 \lesssim 0.33\text{ km s$^{-1}$}$ (99\% CL).
Including the EDGES $T_{21}$-measurements,
we extend our study to constrain the baryon-DM interaction
in the range of DM velocities allowed by our analysis.
As a consequence, we present new bounds on
two electromagnetic models of DM, namely
minicharged particles (MCPs) and electric dipole moment (EDM).
For MCPs, the parameter region that is consistent with EDGES and 
independent bounds on cosmological and stellar physics is very small,
pointing to the sub-eV mass regime of DM.
A window in the MeV–GeV may still be compatible with these bounds
for MCP models without a hidden photon.
But the EDM parameter region consistent with EDGES
is excluded by Big-Bang Nucleosynthesis and Collider Physics.
\end{abstract}


\section{Introduction}
\label{sec:intro}

Within the current cosmological paradigm,
where Dark Matter (DM) dominates in the mass content of the Universe,
the nature of the DM particles plays a key role
in shaping the linear Matter Power Spectrum (MPS) and
the Angular Power Spectrum of the
Cosmic Microwave Background (CMB) anisotropies.
Since the earliest works on the topic,
the empirical evidence has favored collisionless DM particles,
whose velocity dispersion in the early Universe
is so small that perturbations of galaxy size
or larger are not damped by free streaming, i.e., the particles are cold
\citep[][]{Peebles1982,Blumenthal+1984,Davis+1985}.
The Cold DM scenario is actually fully consistent with
current CMB and large-scale structure (LSS) data
\citep[see e.g.,][]{Planck2015,Planck2018cosmpar}.
However, at small scales, this scenario seems to face issues,
especially related to the abundance and properties of dwarf galaxies
\citep[for a recent review, see][]{Bullock2017small}.
Early studies based on N-body cosmological simulations
have shown that these potential issues are alleviated if the DM particles are warm
\citep[][]{Colin+2000,Bode2001,Avila-Reese+2001}.
More recent works, using semi-analytical models and 
N-body + Hydrodynamics cosmological simulations confirm that
the Warm DM scenario for particle masses within a given range,
while keeping the success of the Cold DM one at large scales,
helps to solve their potential issues at small scales
\citep[e.g.,][for more references, see the review by
\citealt{Abazajian2017sterile}]{Lovell+2012,Lovell+2016,Colin+2015,Gonzalez-Samaniego+2016,Bozek+2016,Bose+2017}.

Thus, one of the key pieces of the DM puzzle remains up in the air,
whether it is entirely cold or mildly warm.
Moreover, none of the popular Cold DM candidates
has been detected so far, neither directly nor indirectly.
Consequently, the broad window of DM possibilities is still
open for a rich variety of particles conceived in
extended theories of the Standard Model.
Among the most relevant and general properties of the DM particles are
their rest mass $m_\chi$ and relic velocity dispersion $\Delta\upsilon$.
In this sense, {\it it would be very useful to constrain
these properties in a generic way with the CMB and LSS data.}

On the other hand, the radio signal recently detected by the
Experiment to Detect the Global Epoch of Reionization Signature
\citep[{\sc edges},][]{Bowman2018},
not only represents the first evidence of
the epochs of the Cosmic Dawn,
but its anomalous absorption profile also suggests
the first sign of DM non-gravitational interactions with baryons.
The observed absorption trough was found too deep
compared to previous theoretical notions,
albeit one explanation (among others discussed below)
could be that baryons were cooled down through
some interaction with DM
\citep[see \textit{e.g.}][]{Dvorkin2014,Tashiro2014,%
Munoz2015,Barkana2018possible,Berlin2018,Safarzadeh2018}.
In this regard,
noteworthy studies have included cosmological
data such as the CMB and Lyman-alpha (Ly-$\alpha$) forest,
providing valuable insights into the physics
involving baryon-DM interactions,
especially in the mass regime above MeV's
\citep{Chen2002,Xu2018,Slatyer2018early,Boddy2018First,%
Boddy2018critical,Gluscevic2018,Kovetz2018}.
An important ingredient of this scenario
that has received little attention is
the DM relic velocity dispersion
mainly for particles lighter than a few MeV's,
even though its effects may play a major role.

In this paper,
we explore limits of DM velocity dispersion using LSS and CMB data.
Because for supermassive particles the velocity dispersion
would be irrelevantly small,
we choose to explore a broad range of masses,
from $10^{-3}$ to $10^9$ eV's.
Using these limits and the 21-cm Cosmic Dawn observations,
we propagate the phenomenology to explore
their collisional cross-section with baryons,
taking a velocity dependent interaction of the form
$\sigma\propto v^{-4}$ and $v^{-2}$.
Following these prescriptions, we pay particular attention to
the constraints on the DM minicharge and the electric dipole moment.

For our exploration, we employ the simplest assumption for a generic 
DM momentum distribution, the Boltzmann or Gaussian function.
This is not only simple but arguably the most physically motivated
momentum distribution for DM. For example, thermal relics of
Weakly Interacting Massive Particles (WIMPs)
would obey classical Maxwell-Boltzmann statistics.
On the other hand,
if Axions or other Weakly Interacting Slim Particles (WISPs)
were produced through a non-thermal injection or a phase transition,
they would be described by a narrow Gaussian momentum distribution.

Throughout this paper we adopt the term \textit{velocity dispersion}
as the expectation value $\left\langle p/m \right\rangle$,
weighted with an specific momentum distribution $f(p)$.
For thermal relics, this is known as the \textit{thermal velocity}.
Our focus is on the primordial velocity dispersion of
DM particles through the study of cosmological data
in the linear regime.%
\footnote{On local scales, 
DM dynamics is influenced by gravitational infall,
violent relaxation, and astrophysical feedback effects,
so that, the DM velocity dispersion becomes
much different than the primordial value.
For example, in the Milky Way halo the velocity dispersion 
distribution deviates from the isotropic case,
attaining, e.g., radial velocity dispersion values of
$\sim 200$ km s$^{-1}$ at the maximum \citep{Bird+2019}.}

The rest of this paper is organized as follows:
in section \ref{sec:neutrinos} we briefly review
the types of DM according to their mass and velocity dispersion.
In section \ref{sec:observations}
we constrain the DM velocity dispersion today,
evolving a non-interacting fluid described by a
Gaussian momentum distribution.
Then, in section \ref{sec:21cm} we connect our results with the mechanism
of baryon-DM interactions proposed to cool down the baryonic gas.
In section \ref{sec:em_dm},
the constraints found on the mass, velocity,
and scattering cross-section are then translated to
the DM minicharge and electric dipole moment.
Our conclusions are summarized in section \ref{sec:close}.

\section{Hot, Warm, and Cold DM}
\label{sec:neutrinos}
Before getting in details of our analysis,
it is worth to briefly review general categories of DM.
We do not intend a comprehensive summary
but simply to articulate generic types of DM
according to (not only their mass but) their velocity dispersion.

Cold DM is the most studied type of DM,
for which the free-streaming scale is very small.
Cold DM perturbations above this scale can be modeled as
a perfect fluid with zero pressure or as
a collisionless fluid with zero velocity dispersion.
Beyond the Standard Model theories like Supersymmetry favor 
a large category of Cold DM particle candidates called WIMPs,
with masses 1 GeV $\lesssim m_\chi\lesssim$ 3 TeV,
which have been the target of most indirect and direct detection efforts
\citep[see e.g.,][]{Gaskins2016indirect,Liu2017direct}.
If WIMPs were in thermal equilibrium in the early Universe,
they obeyed a Boltzmann momentum distribution
$\sim e^{-p^2/2M_w T_w}$,
whose associated thermal velocity is
$\left\langle p/M_w \right\rangle = \sqrt{8T_w/\pi M_w}$.
This kind of heavy DM candidates should have decoupled
very early in the radiation dominated era from a cosmic plasma
with a large number of relativistic degrees of freedom (dof)
$\mathrm{g}^*_{w,\text{dec}}$.
From the conservation of the specific entropy,
we know that the WIMPs temperature
is related to the radiation temperature as
$T_w \propto (\mathrm{g}^*_0/\mathrm{g}^*_{w,\text{dec}})^{2/3}\; T_\gamma^2/M$,
where $\mathrm{g}^*_0\sim4$ are the relativistic dof today.
It is pretty clear that for extremely large masses,
the WIMPs temperature (and consequently their thermal velocity)
would be extremely small as well.

Axions are another noteworthy Cold DM candidate.
Originating from the Peccei-Quinn solution to
the strong CP problem \citep{PecceiQuinn1978},
the QCD axion acquires a typical mass of
$\sim 10^{-5}-10^{-2}$ eV,
near the QCD phase transition \citep{Marsh2016axionrev}.
At this time any interaction was already 
suppressed by the Peccei-Quinn scale;
hence, axions would have been produced out-of thermal equilibrium
and thus they are not subject to thermal velocities.
A more general family of axion-like particles
in a broad range of masses ($10^{-24} - 10^3$ eV)
could be produced also non-thermally via the
vacuum misalignment mechanism \citep{Ringwald2012axionrev}.
Furthermore, if by some mechanism
axions are brought into thermal equilibrium,
they would undergo a Bose condensation
\citep[BEC,][]{Sikivie2009bec,Sikivie2012thermal}.
In either case, axions shall be well described by a Boltzmann-like
distribution $\sim e^{-p^2/\Delta p^2}$,
where the momentum width $\Delta p$
(extremely small for axion Cold DM)
encodes the physics of the process leading to the non-thermal state%
\footnote{BECs are commonly referred to as non-thermal states,
albeit their physical origin is obviously thermal.}.

Though being by far sub-dominant,
Hot DM is the best-known component of DM
because it is mainly composed of active neutrinos
\citep{Abazajian2016active}.
Neutrinos were in thermal equilibrium in the early Universe,
obeying the Fermi momentum distribution
in the relativistic limit $( e^{p/T_\nu}+1 )^{-1}$.
Unlike for any other DM candidate,
the decoupling temperature is fairly well known
$T_{\nu,\text{dec}} \approx \text{1 MeV}$
\citep{Lesgourgues2006review}.
At that time only $e^\pm$ and $\gamma$
contributed to the relativistic dof,
$\mathrm{g}^*_{\nu,\text{dec}}=10.75$.
After decoupling, their temperature is proportional to
the one of photons $T_\nu = (4/11)^{1/3}\, T_\gamma$,
and then gets simply red-shifted.
Neutrinos become non-relativistic at late epochs
composing a small fraction of matter today
$\Omega_\nu h^2 = \sum m_\nu /94\text{ eV}$.
The neutrino thermal velocity is
completely parametrized in terms of their mass,
$\left\langle p/m_\nu \right\rangle \approx %
3.15\,T_\nu/m_\nu \approx 150\,(\text{eV}/m_\nu)$ km s$^{-1}$.
The net effect of active neutrinos is
to wash-out the small scale matter fluctuations
and above the free-streaming wavenumber
$k_\text{fs}(z=0) = 0.01 - 0.1 \, h\text{ Mpc}^{-1}$.
For that reason, cosmological observations tightly constrain
the sum of neutrino masses below the eV-scale.

If a small fraction of axions somehow thermalized
\citep[see \textit{e.g.}][]{Archidiacono2013axionhdm},
then they would obey a Bose distribution
$(e^{p/T_a}-1)^{-1}$.
Thermal axions (and any other sub-eV thermal species)
are also Hot DM candidates with a behavior close to active neutrinos.

Warm DM is an interesting intermediate phase,
characterized by slow particles albeit not zero pressure, 
and consequently with non-negligible free-streaming scales.
Sterile neutrinos are Warm DM candidates well motivated
from theory and invoked by some anomalies
in short-baseline oscillation data \citep{Lasserre2014}.
They can mix with active neutrinos but
do not carry weak interactions \citep{ALEPH2006}.
In similarity to active neutrinos,
sterile neutrinos are often assumed 
to decouple thermally while being relativistic,
obeying the Fermi distribution $( e^{p/T_s}+1 )^{-1}$.
In this case, the relic temperature is unknown
but it should be proportional to the photon temperature too,
$T_s=(\mathrm{g}^*_0/\mathrm{g}^*_{s,\text{dec}})^{1/3}\, T_\gamma$,
where $\mathrm{g}^*_0$ are the relativistic dof today.

Studying sterile neutrino mass bounds is a twofold task:
from LSS considerations and from indirect DM searches.
Assuming a specific value for $\mathrm{g}^*_{s,\text{dec}}$
(for example 106.75 in the Standard Model
and twice as much in Supersymmetry),
the thermal velocity and free-streaming scale
become completely specified by the mass $m_s$
and its effects can be constrained with
measurements of the MPS data
(especially through the Ly$\alpha$ forest
for the scales of interest), leading this to
a lower-limit on $m_s$ \citep{Abazajian2017sterile}.
On the other hand,
a fraction of sterile neutrinos is expected to decay rapidly
leading to a source of mono-energetic photons
with energy close to half of its mass.
A hint of such a decay has been prompted by
the discovery of an unidentified emission line at 3.5 keV
in the stacked X-ray spectrum of galaxy clusters and galaxies 
\citep[see][and references therein]{Abazajian2017sterile}.

The majoron \citep[a scalar boson proposed to explain
the \textit{See-Saw} mechanism,][]{Mohapatra1981}
is also a good Warm DM candidate.
With properties and effects similar to sterile neutrinos,
they can be modeled with a thermal Bose momentum distribution
$( e^{p/T_J}-1 )^{-1}$, and a temperature
$T_J=(\mathrm{g}^*_0/\mathrm{g}^*_{J,\text{dec}})^{1/3}\,T_\gamma$.

In general, thermal relics are defined by the relativistic dof
at the moment of their decoupling $\mathrm{g}^*_\text{dec}$.
Correspondingly, the thermal velocity is 
$\left\langle p/m \right\rangle = \sqrt{8T/\pi m}$
in the case of Boltzmann relics,
$3.15 \ T/m$ in the case of fermions, and
$2.7\ T/m$ in the case of bosons.
Evaluated today, the thermal velocity can be expressed
approximately equal for fermions and bosons,
\begin{equation}
 v_\text{th} \approx 0.2 \left(%
    \frac{\mathrm{g}^*_0}{\mathrm{g}^*_\text{dec}}\right)^{1/3}%
 	\frac{\text{1 keV}}{m}%
    \hspace{2.5mm}\text{km s$^{-1}$},
\label{eq:velther1}
\end{equation}
where $T_\text{cmb}=2.72\text{ K}$ is implicit.
An equivalent parametrization –often used for Warm DM–
can be written indicating explicitly the DM abundance
\citep{Hogan2000,Bode2001},
\begin{equation}
 v_\text{th} \approx 0.06 \left(%
             \frac{\Omega_\chi h^2}{g_\chi}\right)^{1/3}
             \left( \frac{\text{1 keV}}{m} \right)^{4/3} %
             \hspace{2.5mm}\text{km s$^{-1}$},
\label{eq:velther2}
\end{equation}
where $g_\chi$ are the DM particle dof.
These two expressions are equivalent and
hold for relativistic Fermi and Bose thermal species.
Similar expressions can be obtained for Boltzmann relics,
just by multiplying Eq. (\ref{eq:velther1}) by $\sqrt{x_d}/2$
and Eq. (2) by $e^{x_d/3}$;
where $x_d\equiv m/T_d$ accounts for the precise time of
DM kinetic decoupling.%
\footnote{%
Elastic scattering with SM species is usually
responsible for keeping DM particles in thermal equilibrium.
In some models, kinetic decoupling might be assumed to occur
at $x_d\approx 1$ \citep{Lesgourgues2013neutrino}.
On the other hand, WIMP co-annihilation numerical studies
suggest that their freeze-out point is $x_f=m/T_f\approx 20-30$
\citep{Roszkowski_2018}.
In general $x_f$ and $x_d$ are separate unknown parameters
but the uncertainty is one-sided because
the freeze-out should typically precede the kinetic decoupling.}

Non-thermal processes, however, are possible and
play a crucial role in Warm DM models.
For example, an important fraction of sterile neutrinos 
could be resonantly  produced \citep[RP,][]{Abazajian2017sterile}.
RP sterile neutrinos are generated with small velocities,
characterized by a sharp distribution peaked at small momenta.
The average momentum reduction is not unique
and depends on the specific mechanism under consideration.
For instance, the Shi–Fuller mechanism \citep{Shi1999PRL}
predicts an average reduction of
$\left<p\right>_{{\rm min}} \sim 0.25\ \left<p\right>_{{\rm thermal}}$
\citep{Laine2008,Boyarsky2009PRL}.
But \citet{Bezrukov2018} proposed a model implying even smaller values.
In any case, the bounds on the sterile neutrino mass
(or equivalently, their velocity dispersion)
become weaker in the case of resonant production
compared to their thermal counterparts.
Another possible source of non-thermal production is
a late decay of heavy particles,
inducing distortions to an otherwise thermal distribution
\citep{Cuoco2005}.
Lastly, Bose condensation \citep{Rodriguez2013}
of at least a fraction of DM particles is
another example of non-thermal processes,
which would relax the current constraints on
the DM mass and velocities.

Actually, it is not the DM mass but more precisely
its velocity dispersion that defines the free-streaming length,
\begin{equation}
    \lambda_\text{fs}(z) = 2\pi\sqrt{\frac{2}{3}}  \frac{\Delta\upsilon(z)}{H(z)},
\end{equation}
which determines the scale below which DM cannot remain gravitationally confined;
or equivalently in Fourier space, the wavenumbers above which 
matter structures are \textit{washed-out} from the MPS.
The comoving free-streaming wavenumber
$k_\text{fs}(z_{\rm nr})=2\pi a(z_{\rm nr})/\lambda_\text{fs}(z_{\rm nr})$
provides a rough approximation to know the $k$'s below which
the free-streaming effects are negligible,
where $z_{\rm nr}$ denotes the time of non-relativistic transition.
For thermal candidates the free-streaming scale depends on
the mass and $\mathrm{g}^*_\text{dec}$
as they are given in equation (\ref{eq:velther1}).
In the case of non-thermal candidates, their velocity dispersion
(and consequently their free-streaming scale)
depends on the specific model of DM production.

Now, irrespective of the precise nature of DM,
their particles will be described by a momentum distribution
denoted by $f(p)$.
Whenever DM interactions are negligible,
$f(p)$ evolves according to the Vlasov equation $df/dt$=0,
whose perturbations in Fourier space read \citep{Ma1995}
\begin{equation}
  \dot{\Psi}-i\frac{q}{\epsilon_q}\Psi = 
  -\left( (\mathbf{k} \cdot \mathbf{\hat{n}}) \dot{\psi}%
  +i\frac{\epsilon_q}{q} (\mathbf{k} \cdot \mathbf{\hat{n}}) \phi\right)%
      \frac{\partial \ln f}{\partial \ln q},
\label{eq:Vlasov}
\end{equation}
where
$q$=$ap$ is the comoving momentum magnitude,
$\mathbf{\hat{n}}$ is the momentum unit vector,
$\mathbf{k}$ is the Fourier wave vector,
and $a$ is the scale factor.
The dynamic variables are the scalar perturbations
$\psi$, $\phi$ to the homogeneous Lemaître-Friedman metric,
a linear statistical perturbation $\Psi$ to $f(p)$,
and the comoving proper energy
$\epsilon_q \equiv (q^{2}+a^2 m_\chi^{2})^{1/2}$.
In general, equation (\ref{eq:Vlasov}) has to be solved
as a Boltzmann-hierarchy of differential equations.
This is the case for Hot and Warm DM, but not for Cold DM.
In the limit $T \rightarrow 0$ or $\Delta p \rightarrow 0$,
one can cut the Boltzmann hierarchy
\citep[see \textit{e.g.,}][]{Dodelson2003,Mo2010}
and the Vlasov equation (\ref{eq:Vlasov}) reduces to 
\begin{eqnarray}
\dot{\delta}_\chi + \theta_\chi = -\dot{h}/2 &\hspace{1cm}&
\dot{\theta}_\chi + H\,\theta_\chi = 0,
\label{eq:cdm}
\end{eqnarray}
where $\delta_\chi = \delta\rho_\chi/\bar{\rho}_\chi$
is the DM fluctuating over-density, and
$\theta_\chi$ is the peculiar velocity.
In synchronous gauge, $\theta_\chi$ is zero
and Cold DM is evolved only through $\delta_\chi$.
The results obtained from this approach
are valid strictly within the linear regime, as such,
$\Delta\upsilon$ is scale-invariant and is interpreted as
the primordial DM velocity dispersion.

Well inside the non-linear regime,
DM particles are subject to violent processes
depending on the scale and local environments
(e.g. gravitational infall, astrophysical feedback effects, etc.),
so that their phase-density can be considerably modified.
Although the latter is not our case of study,
we mention that some interesting inferences have been attempted
by comparing the primordial and coarse-grained DM phase-densities
\citep{Tremaine-Gunn1979,Madsen1991generalized,Hogan2000,Boyarsky2009lowerbound}.

\section{Generic constraints on the DM velocity dispersion}
\label{sec:observations}

The only knowledge about the DM momentum distribution $f_\chi(p)$
is that it must be peaked at low momenta in order to describe
non-relativistic, (almost-)pressureless matter.
We argue that a Gaussian distribution is a good generic description
because it represents a variety of DM scenarios,
from heavy thermal relics to non-thermal distortions,
and even phase-transitions.
We implement the Gaussian momentum distribution 
\begin{equation}
 f_\chi(p) = \frac{n_\chi}{\pi^{3/2} \Delta p^3} \exp\left(-\frac{p^2}{\Delta p^2}\right)
 \label{eq:distr}
\end{equation}
in \textsc{class} \citep{class2,class4},
replacing the default Cold DM with
this non-Cold DM module described by $f_\chi(p)$.
Here $\Delta p$ is the momentum width and
$n_\chi=\int d^3p f_\chi(p)$ is the number density.
Without regard to the (thermal or not) origin of DM,
we can define a fiducial `temperature'
$T_\chi = \Delta p^2/2m_\chi$,
in terms of which, we can write the velocity dispersion
$\Delta\upsilon = \left\langle p/m_\chi \right\rangle =%
\sqrt{8T_\chi/\pi m_\chi} = 1.13\,\Delta p/m_\chi$.
Notice that the velocity dispersion gets linearly red-shifted
$\Delta\upsilon = \Delta\upsilon_0\,(1+z)$,
being $\Delta\upsilon_0$ the value measured today.

\begin{figure}[t]
 \centering
 \includegraphics[width=\linewidth]{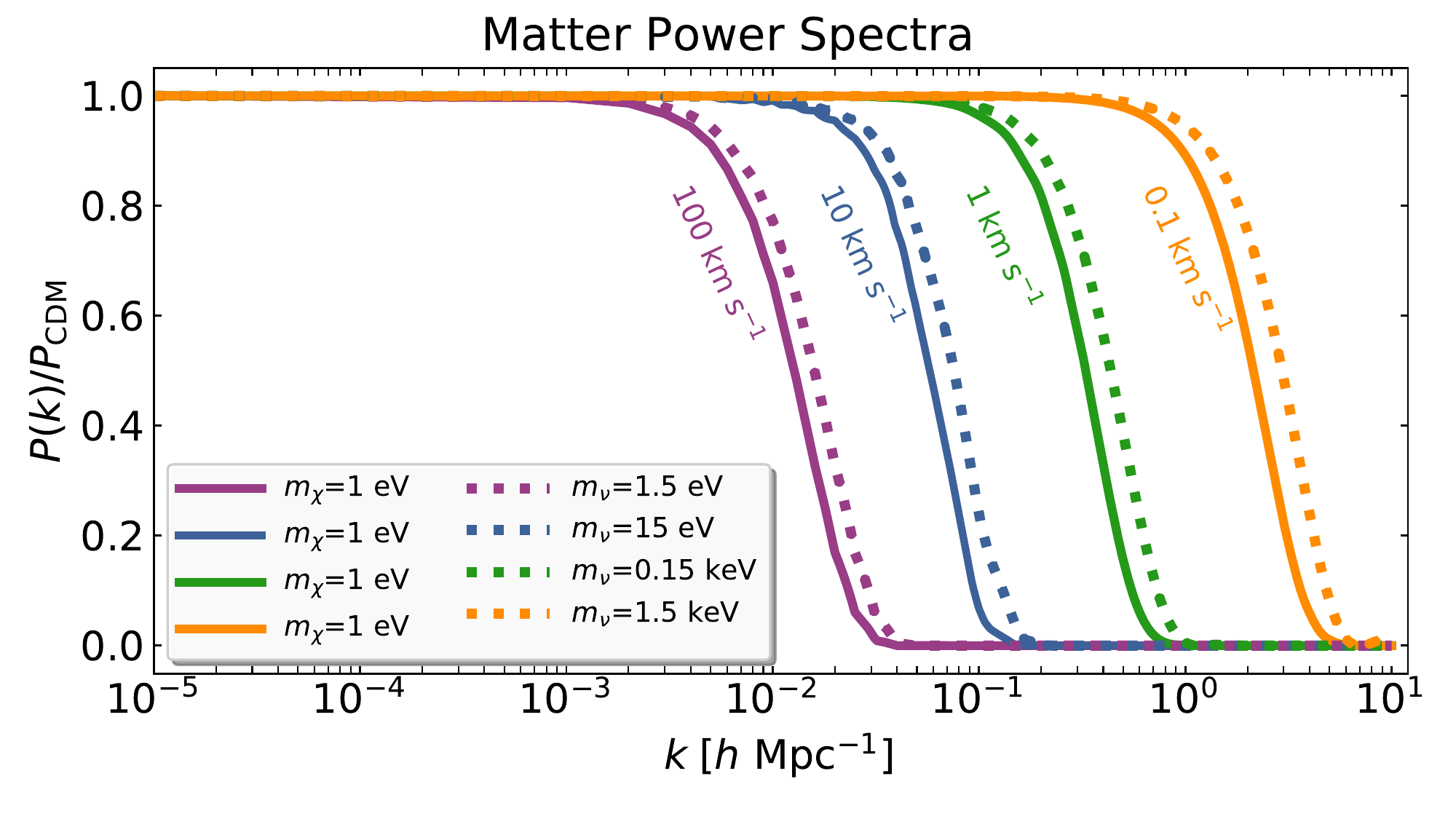}
 \caption{Distinct ratios of the MPS with respect to Cold DM.
    Same color indicate same velocity dispersion.
    Solid lines correspond to 1 eV-mass DM particles described
    by the Gaussian distribution $f_\chi$,
    while dotted lines are thermal neutrinos described by
    the Fermi distribution.
    The neutrino thermal velocity is uniquely specified by its mass.
    In contrast, DM particles described with $f_\chi$
    approach to Cold DM as $\Delta\upsilon_0 \rightarrow 0$,
    irrespective of their mass.
    } 
\label{fig:class_deltap}
\end{figure}

As mentioned in \S\ref{sec:neutrinos},
the DM description with equation (\ref{eq:distr})
reduces to standard Cold DM in the limit $\Delta\upsilon \rightarrow 0$.
This is reproduced in Figure \ref{fig:class_deltap},
where we plot the MPS ratio (over Cold DM)
for fixed $m_\chi$=1 eV but different values of $\Delta\upsilon_0$.
These ratios progressively approach to 1
for smaller values of $\Delta\upsilon_0$.
An analogous effect is produced by a Fermi distribution
that progressively approaches to Cold DM for masses in the keV-range.
But we recall that the velocity dispersion (not the mass) regulates
the free-streaming scale for a given DM model.
In Figure \ref{fig:class_deltap} we show both Gaussian and Fermi
cases for equivalent velocity dispersions,
from which we notice that the Gaussian distribution causes
slightly more suppression on the MPS than Fermi.
Thus, we are showing that the Gaussian distribution $f_\chi(p)$
is a convenient description for DM because through its parameters
it can cover hot, warm, and cold possible states of DM.

Our primary goal is to obtain observational constrains for
$\Delta\upsilon_0$ using public data surveys such as
Planck \citep{planck_temperature_data,planck_lensing_data},
Baryonic Acoustic Oscillations \citep[BAO,][]{bao_boss_dr12,bao_boss_2da},
and Red Luminous Galaxies from the
Sloan Digital Sky Survey \citep[SDSS,][]{sdss_dr4}.
We employ \textsc{montepython} \citep{Montepython2013,Brinckmann2018} to perform
Bayesian estimations using $0 < \Delta\upsilon_0/\text{km s$^{-1}$} < 30$ as a prior.
Although we have seen that the CMB and MPS are insensitive to the DM mass
(when $\Delta\upsilon_0$ is varied independently), 
we choose to check for any marginal effect by splitting the analysis
into six stages from sub-eV to GeV as indicated in Table \ref{tab:priors_results}.
For the rest of the cosmological parameters, we use customary flat priors.
Additionally, for any pair values of $\Omega_\chi h^2$ and $m_\chi$,
$n_\chi$ gets internally rewritten by CLASS in order to satisfy
the equation $\Omega_\chi = m_\chi \, n_\chi/\rho_c$
(with $\rho_c$ the critical density).

After a deep exploration, the standard cosmological parameters
are constrained in concordance with standard reports \citep{Planck2015}.
We find no significant degeneracies between $\Delta\upsilon_0$
and the standard cosmological parameters,
suggesting an independent effect from our parametrization.%
As expected, the mass parameter $m_\chi$ is unconstrained when
the velocity dispersion is varied independently.
As it can be read from Table \ref{tab:priors_results}
and figure \ref{fig:mass_vel_disp},
the constraints on $\Delta\upsilon_0$ are not significantly different
by comparing the six stages of mass-sampling.
The Planck and BAO BOSS data constrain the free-streaming effects
that would be caused by a large DM velocity dispersion
(similarly to an increase on the effective number of relativistic species).
But the most restrictive constraints on $\Delta\upsilon_0$
are obtained when the LSS data are included, 
because the free-streaming suppresses the MPS on small scales.
Summarizing the results, our analysis shows that cosmological data
constrain the DM velocity dispersion to
\begin{eqnarray}
\Delta\upsilon_0 \lesssim 0.33 \text{ km s$^{-1}$}
         && \hspace{0.5cm} \text{(99\% CL)}.  
\label{eq:bounds}
\end{eqnarray}
This translates to a lower bound on the epoch of 
DM non-relativistic transition, $z_{\rm nr}\gtrsim 10^6$.
But because $k \lesssim \text{0.2 h Mpc}^{-1}$ wavenumbers
correspond to modes that entered the horizon at redshifts $z \lesssim 10^5$,
the free-streaming effects of a DM species with 
$\Delta\upsilon_0\ll\text{0.33 km s}^{-1}$
would not be noticeable at the scales of the LSS data used in our analysis.

\begingroup
\setlength{\tabcolsep}{6pt} 
\renewcommand{\arraystretch}{0.8} 
\begin{deluxetable*}{cccccccc}
\tablecaption{\centering Bounds on the Dark Matter velocity dispersion
from the six mass-sampling stages of the analysis.}
\tablehead{  \colhead{} & & Stage 1 & Stage 2 & Stage 3 & Stage 4 & Stage 5 & Stage 6 \\
\colhead{} & & $10^{-3}-10^{-1}$ eV's & $10^{-1}-10^1$ eV's & $10^1-10^3$ eV's & $10^3-10^5$ eV's & $10^5-10^7$ eV's & $10^7-10^9$ eV's}
\startdata 
P   &  & 1.32 & 1.31 & 1.33 & 1.34 & 1.31 & 1.33 \\ 
PB  & $\Delta\upsilon_0$ [km s$^{-1}$]  $\lesssim $ & 1.28 & 1.27 & 1.26 & 1.25 & 1.26  & 1.25 \\ 
PBS & & 0.32 & 0.32 & 0.32 & 0.33 & 0.32 & 0.33 \\
\enddata
\tablecomments{Upper limits at 99\% CL.
The prior on the velocity dispersion is
$0 < \Delta\upsilon_0/\text{km s$^{-1}$} < 30$ for every stage of the analysis.
The datasets are denoted with \textit{P}: Planck, \textit{PB}: Planck + BAO BOSS,
\textit{PBS}: Planck + BAO BOSS + SDSS DR4 LRG.}
\label{tab:priors_results}
\end{deluxetable*}

\begin{figure}[b]
 \centering
 \includegraphics[width=0.5\textwidth]{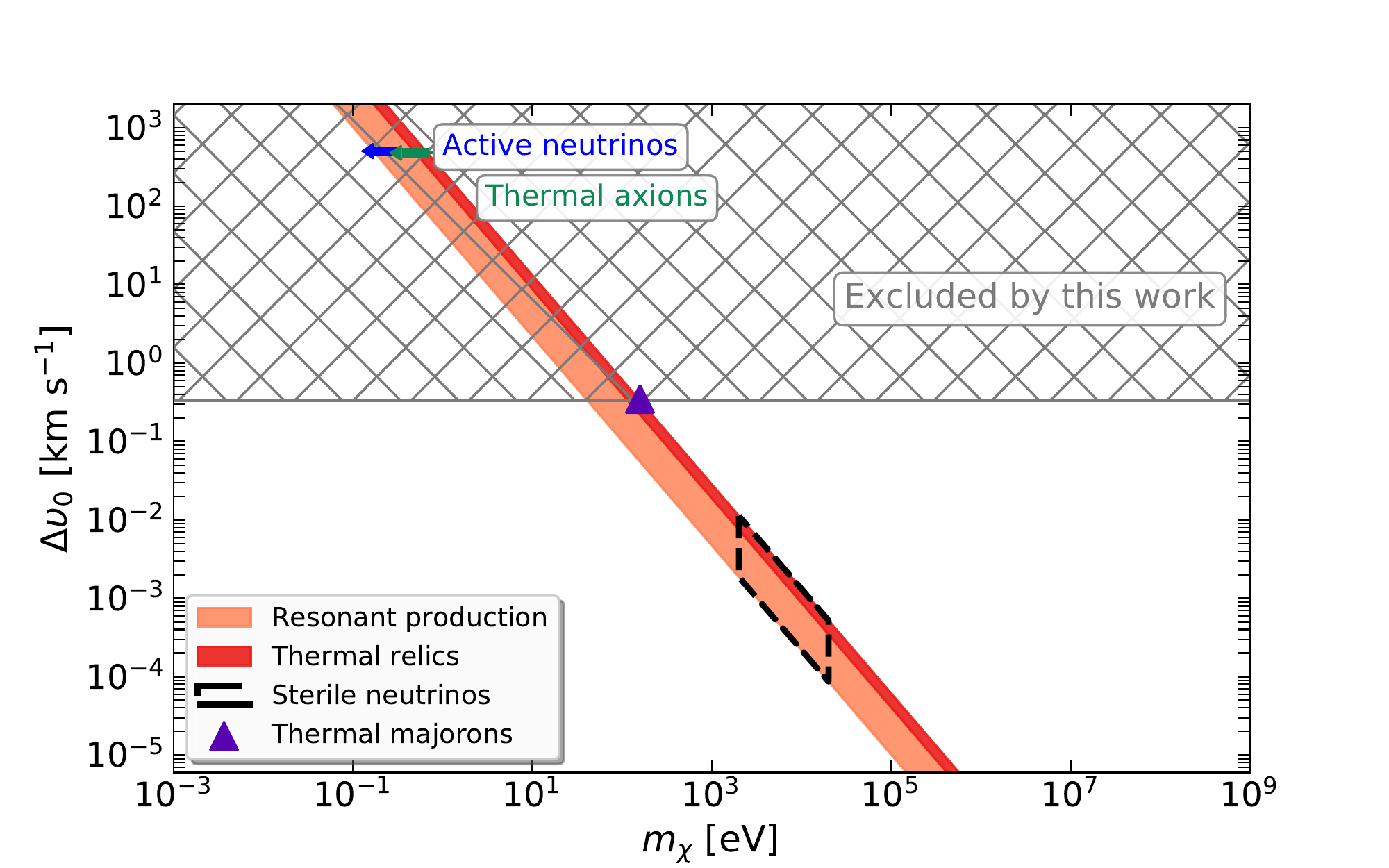}
 \caption{Constraints on the DM velocity dispersion
    after our analysis of the CMB and LSS data.
    The dark-red line shows thermal velocities for Warm DM
    using equation (\ref{eq:velther2}),
    while the light-red band is a 0.25 non-thermal correction
    (due to resonant production in the case of sterile neutrinos).
    Previous reports on sterile neutrino and thermal majorons
    are shown as candidates of Warm DM.
    Reports on Hot DM candidates such as active neutrinos
    and thermal axions are shown using equation (\ref{eq:velther1}).}
\label{fig:mass_vel_disp}
\end{figure}

In figure \ref{fig:mass_vel_disp} we also mark
the thermal velocity, using equation (\ref{eq:velther1}),
due to active neutrinos and thermal axions;
according to previous reports on their masses,
$\sum m_\nu \lesssim 0.23 \text{ eV}$ \citep{Planck2015}
and $m_{a,\text{th}} \lesssim \text{0.67 eV}$
\citep{Archidiacono2013axionhdm}.
We also plot the thermal velocity given in equation (\ref{eq:velther2}),
with $\Omega_\chi\,h^2=0.12$, and $\mathrm{g}_\chi=2$ (dark-red line).
We should correct $v_\text{th}$ in two ways:
First, we have to take into account the uncertainties on the mass,
which is the most unknown parameter,
leading us to the error propagation
$\delta v_\text{th} \approx 4/3 \; v_\text{th} \; (\delta m/m$).
Unfortunately, we do not count on any measurement of the DM mass;
thus, we adopt a conservative choice for the mass error $\delta m= 0.3\,m$ 
(roughly what it might be expected after a preliminary
and speculative evidence of this parameter)
that may help us for illustrative purposes.
Second, we include a Shi-Fuller correction to relax the bounds
in the case of resonant production of sterile neutrinos (light-red band).
Additionally, we ought to account for the uncertainty on the
details of kinetic decoupling
(\textit{e.g.} the value of $\mathrm{g}^*_\text{dec}$
or $x_d$, see text below Eq. \ref{eq:velther1}),
but for illustrative purposes, we just depict the thermal velocity
as it is shown in equation (\ref{eq:velther2}).
We also spot the mass constraint
$m_J = 0.158\pm 0.007 \ \text{keV}$
reported for thermal majorons as
a Warm DM candidate \citep{Lattanzi2013}.
Thermally produced sterile neutrino bounds are still in debate,
there is a controversy between lower bounds from cosmological data
and upper bounds from diffuse X-ray emission
\citep[see these comprehensive reviews][]{Adhikari2017white,Boyarsky2018review}.
In contrast to thermal relics, resonantly produced (RP)
sterile neutrinos do not need large mixing angles
with active neutrinos to match the required DM abundance.
As a consequence, RP sterile neutrino decays into
X-rays may be suppressed, thereby loosening the mass upper bound.
Recent analyses of Ly-$\alpha$ forest data set a lower limit
$m_s \gtrsim \text{5.3 keV}$ on thermally produced sterile neutrinos 
\citep[][which is larger than previous determinations]{Irsic2017}.%
\footnote{Notice that these constraints are still subject
to uncertainties on the thermal evolution
of the Intergalactic Medium.}
Meanwhile, combined analyses of SDSS/BOSS and Ly-$\alpha$ forest data
set a lower limit $m_s \gtrsim \text{3.5 keV}$
on RP sterile neutrinos \citep{Baur2017rpsn}.
A noteworthy recent report based solely
on the EDGES signal measured timing, sets
a lower limit of $m_s \gtrsim \text{2 keV}$
\citep{Safarzadeh2018}.
On the other hand, the non-observation of X-ray photons 
induced by the decay of sterile neutrinos
sets an upper bound of $m_s \lesssim \text{ 20 keV}$
\citep{Adhikari2017white}.
All in all, we include in figure \ref{fig:mass_vel_disp}
(see black-dashed lines)
mass constraints on both thermal and RP sterile neutrinos
within $\text{2 keV} < m_s < \text{20 keV}$,
corresponding to a velocity dispersion within
$5\times 10^{-5}$ km s$^{-1}$ $\lesssim v^s_0 \lesssim$ $10^{-2}$ km s$^{-1}$.

Figure \ref{fig:mass_vel_disp} displays a wide region
of allowed $m_\chi$ and $\Delta\upsilon_0$ parameters;
let us now place our bounds in context.
We begin to recall that active neutrinos and thermal axions
–as well as any other Hot DM species–
are clearly discarded as the main source of DM.
In the case of thermal DM (or non-thermal one with a correction of 0.25)
keep in mind that the respective dark and light red bands bear
large uncertainties in the mass
and kinetic decoupling parameters,
so we use them only for illustrative purposes.
With that in mind, we might read that our bounds seem 
to disfavour thermal candidates 
(including non-thermal corrections) with masses below $\lesssim 40$ eV.
A previous report on thermal majorons \citep{Lattanzi2013}
is at the edge but within of our 99\% CL bounds.
Previous reports on thermal \citep{Irsic2017} and 
resonant \citep{Baur2017rpsn} sterile neutrinos
are well below our 99\% CL boundary.
Notice again that our bounds do not exclude any DM mass
from $10^{-3}$ to $10^9$ eV's.
Indeed, Warm ($0 < \Delta\upsilon_0 \lesssim 0.33 \text{ km s}^{-1}$)
and Cold ($\Delta\upsilon_0 = 0$) DM candidates 
are well allowed by our constraints, irrespective of their mass.

\section{Role of DM velocity dispersion on the 21-cm signal}
\label{sec:21cm}
In many studies of the 21-cm Cosmology,
it is customary to fix the DM mass to the WIMPs scale,
whose corresponding thermal velocity is nearly zero.
But from the previous section we see that a vast variety
of DM candidates could involve a significant velocity dispersion
while still reproducing the observed LSS and CMB spectra.
Now we are going to use the constraints of Figure
\ref{fig:mass_vel_disp} in order to explore
the $\Delta\upsilon_0$ effects
on the interpretation of the EDGES measurements.

EDGES probes the epochs after primordial recombination and
before the formation of the first luminous sources.
During these epochs, the baryonic gas is mainly composed of
neutral hydrogen with total spin S=0
(proton/electron anti-parallel spins).
When an atom in the parallel state (S=1) realigns its spins,
a photon is emitted with an energy $E_{21}$=5.87 $\mu$eV, 
equivalent to a wavelength of 21 cm.
The 21-cm signal is the observed brightness temperature
with respect to the photon background
\citep[for a comprehensive review see:][]%
{Furlanetto2006,Morales2010,Pritchard-Loeb2012},
\begin{equation}
T_{21}(z) = (27\text{ mK})\frac{x_{HI}\Omega_b h^2}{0.023}
         \left(1-\frac{T_\gamma(z)}{T_s(z)}\right)
         \left(\frac{0.15}{\Omega_mh^2}\frac{1+z}{10} \right)^{1/2} .
\label{eq:T21}
\end{equation}
Here, $x_\text{HI}$ ($\approx$1 during the epoch of cosmic dawn)
is the fraction of neutral hydrogen
and $\Omega_b$ is the baryon abundance.
$T_s$ is called the `spin temperature',
which defines the relative population of the two spin levels
$n_1/n_2 \equiv 3 e^{-E_{21}/T_s}$,
it can be parametrized in terms of 
the baryon and photon temperatures, $T_b(z)$ and $T_{\gamma}(z)$,
and the stimulated Ly-$\alpha$ emission \citep{Chen2004}.
In the limit of full Ly-$\alpha$ coupling,
we take $T_s=T_b$ \citep{Madau1997}.
The 21-cm signal is then redshifted
till the band of radio-frequency today.
The EDGES collaboration reported
$T_{21}=-0.5^{+0.2}_{-0.5}$ K (99\% C.L.)
in a redshift range 13 $\lesssim z \lesssim$ 22,
centered at $z\approx 17$
(or a frequency of 78 MHz).

Some aspects of the reported absorption profile
are peculiar and need to be explained:
the early redshift range with their implications for
star formation \citep{Madau2018,Mirocha2018what},
the flat shape of the profile \citep{Venumadhav2018heatingIGM},
and the unanticipated deep trough.
Assuming only standard physical scenarios,
the maximum value of the absorption trough would be
$T_{21}\approx -0.2$ K, \textit{i.e.}
the measurement is at least twice the standard expectation.
Given that $T_{21}$ depends on the ratio $T_\gamma/T_s$,
two main explanations are currently discussed to enhance the absorption:
\textit{i}) An excess of radiation injected from
DM-annihilations, black holes, or any other astrophysical source
\citep[see \textit{e.g.}][]{Chianese2018,Clark2018,Feng2018enhanced,
Sharma2018}.
\textit{ii}) A cooling mechanism of baryons through interactions with DM
\citep[see \textit{e.g.}][]{Dvorkin2014,Tashiro2014,Munoz2015,
Barkana2018possible,Berlin2018,Slatyer2018early,Xu2018}.
It should also be mentioned that the EDGES findings
are being argued to be due to systematics
related to residual foregrounds \citep{Hills2018concerns}.
Thus, the EDGES observations require confirmation 
from similar experiments like
SCI-HI \citep{SciHi2014},
LEDA \citep{LEDA2016}, and
SARAS 2 \citep{SARAS2017}.
Ultimately, the EDGES results open a rich discussion
pointing to new physics and novel phenomenological frameworks.

Here we focus on the baryon-DM interaction hypothesis,
assuming a velocity-dependent scattering cross-section
$\sigma(v)=\sigma_0 \, v^n$, where
$v=|\mathbf{v_\chi} - \mathbf{v_b}|$
is the relative velocity between two particles%
\footnote{It is also customary to use a parameter $\sigma_1$
that relates to $\sigma_0$ as
$\sigma_0 = ([1\text{ km s$^{-1}$}]/c)^4\, \sigma_1$.}.
We choose to explore two cases of low-energy enhanced interactions,
namely $n=-4$ and $n=-2$,
which are motivated by models of
minicharge and electric dipole moment, respectively.
Other cases of $n$ have been studied elsewhere,
\citep[see \textit{e.g.},][]{Dvorkin2014,Slatyer2018early,Xu2018}.

The thermal evolution of baryons and DM involves
the baryon and DM temperatures, $T_b(z)$ and $T_\chi(z)$;
the energy transfer between baryons and DM, $Q_b$ and $Q_\chi$;
and the relative bulk velocity $V_{\chi b}$.
The full formalism can be read up on
\citet{Dvorkin2014}, \citet{Tashiro2014}, or \citet{Munoz2015};
let us just discuss the baryon-DM energy transfer,
\begin{equation}
 \dot{Q}^{(n)}_{b} = \sum_{t=e,p} \frac{ f_{{\rm dm}}\rho_{\chi} \sigma_0 \, m_t}%
                               {(m_t + m_\chi)^2}%
 \left( \frac{(T_{\chi}- T_b)S_n(r_t)}{u_t^{-    (n+1)}}
             + \frac{m_{\chi}F_n(r_t) }{V^{-(n+3)}_{\chi b}}\right),
\label{eq:heat-transfer}
\end{equation}
where $t$ refers to the proton or electron,
$f_\text{dm}$ is the fraction of DM interacting with baryons,
$u_t = \sqrt{T_b/m_t + T_\chi/m_\chi}$
is the thermal width of the relative bulk velocity,
and $r_t=V_{\chi b}/u_t$.
Notice that the first term subtracts energy
from baryons as long as $T_\chi$ < $T_b$.
This cooling term is suppressed by the functions $S_n(r_t)$,
where $S_{-4}(r_t)=\sqrt{2}e^{-r_t^2 /2}/\sqrt{\pi}$ and
$S_{-2}(r_t)=2{\rm Erf}(r_t/\sqrt{2})/r_t$.
The second term transforms the mechanical energy
into heating to both baryons and DM, where
$F_{-4}(r_t)={\rm Erf}(r_t/\sqrt{2})-\sqrt{2}r_te^{-r_t^2/2}/\sqrt{\pi}$
and 
$F_{-2}(r_t)= {\rm Erf}(r_t/\sqrt{2})-r_t^{-2}F_{-4}(r_t)$.
This term can spoil the cooling mechanism unless
the velocity ratio $r_t$ is small ($V_{\chi b} \ll u_t$),
or the DM mass is small ($ m_\chi \ll m_t$).

The hypothesis of baryon-DM scattering
has been extensively studied using CMB and MPS observables
(the same observables that we just used
to constrain $\Delta\upsilon_0$),
which are made of modes that entered the horizon at $z\sim 10^3-10^5$
(with the Ly-$\alpha$ forest it is possible to probe beyond $z\gtrsim 10^6$).
At those early times,
the overall effect of baryon-DM interactions
would mimic an increase of the baryonic budget;
consequently, both the CMB and MPS would become damped
on small scales \citep{Chen2002,Dvorkin2014}.
This effect is actually a very good probe for
the strength of baryon-DM scattering
and previous studies have reported tight upper limits:
$\sigma_0 \lesssim 10^{-41} \text{ cm}^2$ for $n=-4$
\citep{Slatyer2018early,Xu2018,Boddy2018critical}, and
$\sigma_0 \lesssim 10^{-33} \text{ cm}^2$ for $n=-2$
\citep{Xu2018,Boddy2018critical}.

On the other hand,
the precise fraction of interacting DM is still in debate
\citep[see \textit{e.g.}][]{Dolgov2013millicharge,Dolgov2017millicharge}.
According to constraints derived from the CMB,
$f_\text{dm}$ might be expected below
the fractional uncertainty of the baryon energy density
\citep{Kovetz2018tighter}.
Besides, in order to avoid other astrophysical constraints,
a small $f_\text{dm}$ might be necessary as well
\citep{Chuzhoy2009minicharged,McDermott2011}.
In the following discussion we consider
$f_\text{dm}= \{1,0.1,0.01\}$
only for exploratory purposes.

Taking into account the smallness of
$\sigma_0$ and $f_\text{dm}$ from linear Cosmology,
the baryon-DM interactions would have little impact
on the distribution of DM velocities
and its dispersion should evolve as
$\Delta\upsilon (z) = (1+z)\,\Delta\upsilon_0$
right after photon decoupling.
Afterwards ($z\lesssim 10^3$), the low-velocity enhanced
scattering ($n=-4$ or $n=-2$)
cause a late-time coupling between DM and baryons.
Although the non-interacting fraction will preserve an
adiabatic dilution ($\Delta\upsilon_\text{\tiny NI}\sim 1+z$),
the interacting fraction will be heated and
its velocity dispersion $\Delta\upsilon_\text{\tiny I}(z)$
will not follow a linear evolution with $z$.
In fact, $\Delta\upsilon_\text{\tiny I}(z)$
can only be computed numerically after the solution to the
heat transfer equations involving (\ref{eq:heat-transfer}).
Thus, when interactions are effective,
the average DM velocity dispersion is
\begin{equation}
    \Delta\upsilon_\text{av}(z) = f_\text{dm}\,\Delta\upsilon_\text{\tiny I}(z) + (1-f_\text{dm}) \, \Delta\upsilon_\text{\tiny NI}(z).
\end{equation}
Given that the DM velocity dispersion is constrained
directly by the effective amount of matter needed for LSS formation,
the constraints found in the previous section should
approximately hold even in the interacting case.
Indeed, the upper-limit in equation (\ref{eq:bounds})
can be used as an initial condition,
\begin{eqnarray}
    \Delta\upsilon_\text{av}(z) \lesssim (1+z)\,\Delta\upsilon_0,
    && \hspace{0.5cm}
    \text{at $z\approx 10^3$}.
    \label{eq:interacting_bound}
\end{eqnarray}
Now, in order to solve the 21-cm thermal dynamics,
we can substitute
$T_\chi = (\pi/8)\,m_\chi\, \Delta\upsilon_\text{av}^2$
in the thermal width $u_t$,
starting the integration at an epoch
much before the Cosmic Dawn ($z\sim 10^3$),
and use equations (\ref{eq:interacting_bound}) and (\ref{eq:bounds})
to define a set of initial conditions.

Let us intuitively discuss the kinematics involved in
the 21-cm thermal evolution and heating transfer.
We already mentioned below equation (\ref{eq:heat-transfer})
that there is a competition between
the cooling mechanism (first term)
and the mechanical heating (second term).
The most obvious way to enhance the baryon cooling is having a large
$\sigma_0$, albeit possibly conflicting with cosmological bounds.
Also obvious is the fact that the colder DM initially is,
the easier for it to absorb heat from baryons;
this is easily seen in equation (\ref{eq:heat-transfer})
because a smaller thermal width $u_t$ enhances the cooling term.
Contrarily, the mechanical heating can overcome
the cooling mechanism in some cases.
For example, particles as heavy as 1-10 GeV would need
$\sigma_0 \gtrsim 10^{-39} \text{ cm}^2$
in order to explain the EDGES signal,
even in the case of $f_\text{dm}=1$ and zero initial
DM velocity dispersion \citep{Barkana2018possible};
clearly conflicting with cosmological bounds.
The mechanical heating can be suppressed though,
if the DM mass is small enough ($ m_\chi \ll m_t$) and/or
the velocity ratio $r_t$ is small ($V_{\chi b} \ll u_t$).
The latter has also been identified as a necessary condition
to maintain linearity in the perturvative Boltzmann equations
\citep[see e.g.][]{Boddy2018First,Slatyer2018early,%
Kovetz2018tighter,Boddy2018critical}.

\begin{figure}[t]
 \centering
 \includegraphics[width=0.48\textwidth]{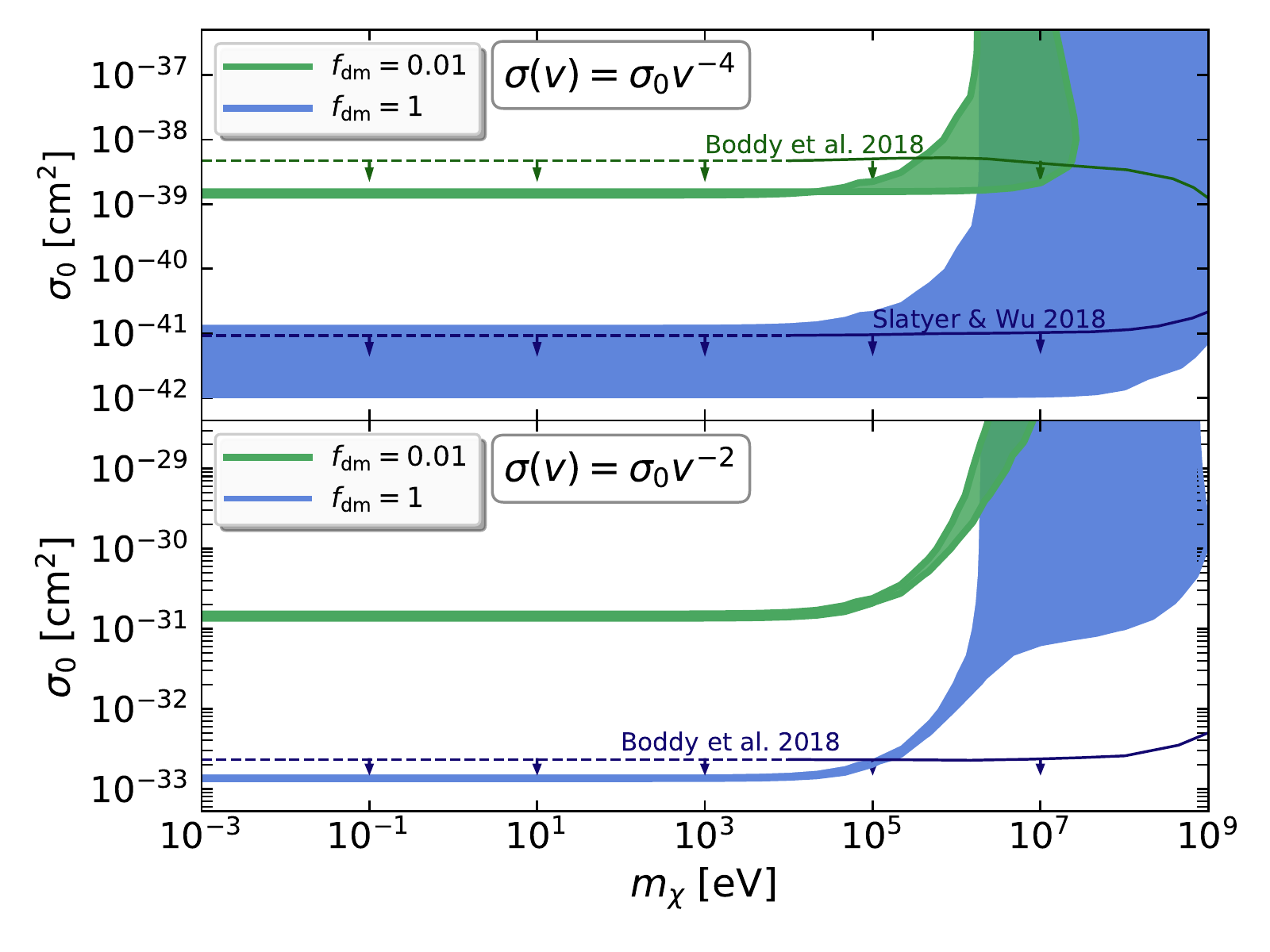}
 \caption{Constraints on the baryon-DM scattering cross-section
     required to explain EDGES signal,
     reported in terms of the mass and marginalized over
     $\Delta\upsilon_0$.
    The blue (green) region represents the 99\% CL
     for $f_{{\rm dm}}=1.0$ ($f_{{\rm dm}}=0.01$).
     Each solid line represents the Planck upper bounds on the cross section from \citep{Slatyer2018early} and \citep{Boddy2018critical}
     (blue for $f_{{\rm dm}}=1.0$ and green for $f_{{\rm dm}}=0.01$);
     dashed lines are an extrapolation to smaller DM masses.
    }
\label{fig:sigma_m}
\end{figure}

The DM mass is a very interesting parameter in this framework.
Given that the absorbed heat is distributed among
the number of interacting DM particles,
the cooling mechanism seems to be easier if the DM mass is small.
In other words, the lighter DM is,
the transferred energy is spread out over more particles,
which could be understood as a more efficient
thermal reservoir than in the heavier case.

We also identify a couple of differences between the two types of scattering:
\textit{i}) For $n=-4$,
the electron-DM interaction is only relevant if DM particles are not cold,
otherwise $\sigma_0$ is dominated by proton-DM interactions.
\textit{ii}) On the other hand, for $n=-2$,
the scattering is electron dominated for
$m_\chi\lesssim 5\times 10^5\text{ eV}$ (even for Cold DM particles);
above that mass, the interaction is proton dominated.

Now we can proceed to fit the 21-cm temperature
appearing in equation (\ref{eq:T21}) to the EDGES measurement.
In practice, for each pair of $\Delta\upsilon_0$, $m_\chi$ fixed values,
we solve for $\sigma_0$ to recover $T_{21}(z=17)\approx-0.5$ K.
Figure \ref{fig:sigma_m} displays the resulting allowed regions for
$\sigma_0$ and $\,m_\chi$ marginalized over $\Delta\upsilon_0$,
for $n=-4$ and $n=-2$,
and the fractions $f_\text{dm}$=1 and 0.01.
Notice that if the initial thermal width $u_t$ is large,
it will suppress the cooling term in equation (\ref{eq:heat-transfer});
such a suppression can only be compensated by $\sigma_0$,
requiring stronger baryon-DM interactions.
The larger values of $\sigma_0$ in Figure \ref{fig:sigma_m}
are clearly in conflict with typical cosmological bounds%
\footnote{Yet, recall that current cosmological bounds
have been mainly focused on DM masses above  MeV's.}.
Indeed, with a tighter bound on $\Delta\upsilon_0$ resulting
from small-scale LSS data like \textit{e.g.} Ly-$\alpha$ forest,
the allowed space for $\sigma_0$ would shrink below the current cosmological bounds.
For masses above 0.1 MeV's, the explanation of the EDGES measurement requires
heavy DM particles to be initially very cold.
But lighter DM particles are much less restricted
on their velocity dispersion initial conditions,
which is due to the aforementioned better cooling efficiency of light-DM particles.

\section{Electromagnetic properties of DM?}
\label{sec:em_dm}
In this section we discus the physical motivation
for the $n=-4$ and $n=-2$ scattering cases,
relating them to the electric minicharge $\epsilon$
and the electric dipole moment $\mathcal{D}$ of DM, respectively.

\subsection{Minicharge}
\label{sec:mcp}
The possible existence of new particles endowed with a small electric charge
$q_\chi=\epsilon e$ (with $e$ the electron charge and $\epsilon \ll 1$)
is well motivated from simple extensions of the Standard Model
that include a hidden sector with an U'(1) unbroken gauge symmetry
\citep{Holdom,Foot}.
The small effective charge is a byproduct of the kinetic mixing
between hidden photons associated with U'(1) and ordinary photons.
Then, fermions in the hidden sector charged under U'(1) can
couple to ordinary photons via $q_\chi$.
If there were light charged scalars in the hidden Higgs sector, 
they would also acquire a tiny charge $q_\chi$ due to the
photon mixing \citep{Melchiorri2007minicharge,Ahlers2008,An2013}.
In some models, even neutrinos are explicitly allowed
to acquire a small charge \citep{Foot,Vinyoles2016}.
It turns out quite intuitive to think of Minicharged Particles (MCPs)
to account for at least a fraction of the DM \citep{Goldberg}.
Indeed, MCPs are often quoted within the group of WISP-DM candidates
\citep{Jaeckel2010,Ringwald2012axionrev}, including
dark photons, majorons, axions, and axion-like particles.

The longstanding question about MCPs
has led to several laboratory searches,
like the experiments at the SLAC National Accelerator Laboratory,
uniquely designed to detect MCPs
\citep{Prinz1998SLAC,Badertscher2007,Gninenko2007,Batell2014SLAC}
that have set an upper bound 
$\epsilon \lesssim 10^{-5}$ in the 0.1 to 100 MeV mass range.
Collider precision tests have set bounds going down to 
$\epsilon \lesssim 5\times 10^{-4}$
\citep{Davidson2000} for masses below 100~keV.
Meanwhile, astrophysical and cosmological environments represent
advantageous laboratories as many of them
are sensitive to the effects of MCPs.
For instance, Big Bang Nucleosynthesis (BBN) sets the condition
$\epsilon \lesssim 10^{-8}$ \citep{Mohapatra1990}
in order to prevent late thermalization of $\lesssim$ MeV particles.
Otherwise, DM would contribute with extra relativistic dof,
which are tightly constrained to
$N_\text{eff}=2.94 \pm 0.38$ \citep{BBN2016}.
If one counts the extra relativistic dof due to the hidden photons,
then, the CMB bounds on $N_\text{eff}$ also place constraints
on the MCP parameter space
\citep[see \textit{e.g.},][]{Vinyoles2016,Barkana2018Strong}.

\begin{figure}
 \centering
 \includegraphics[width=\linewidth]{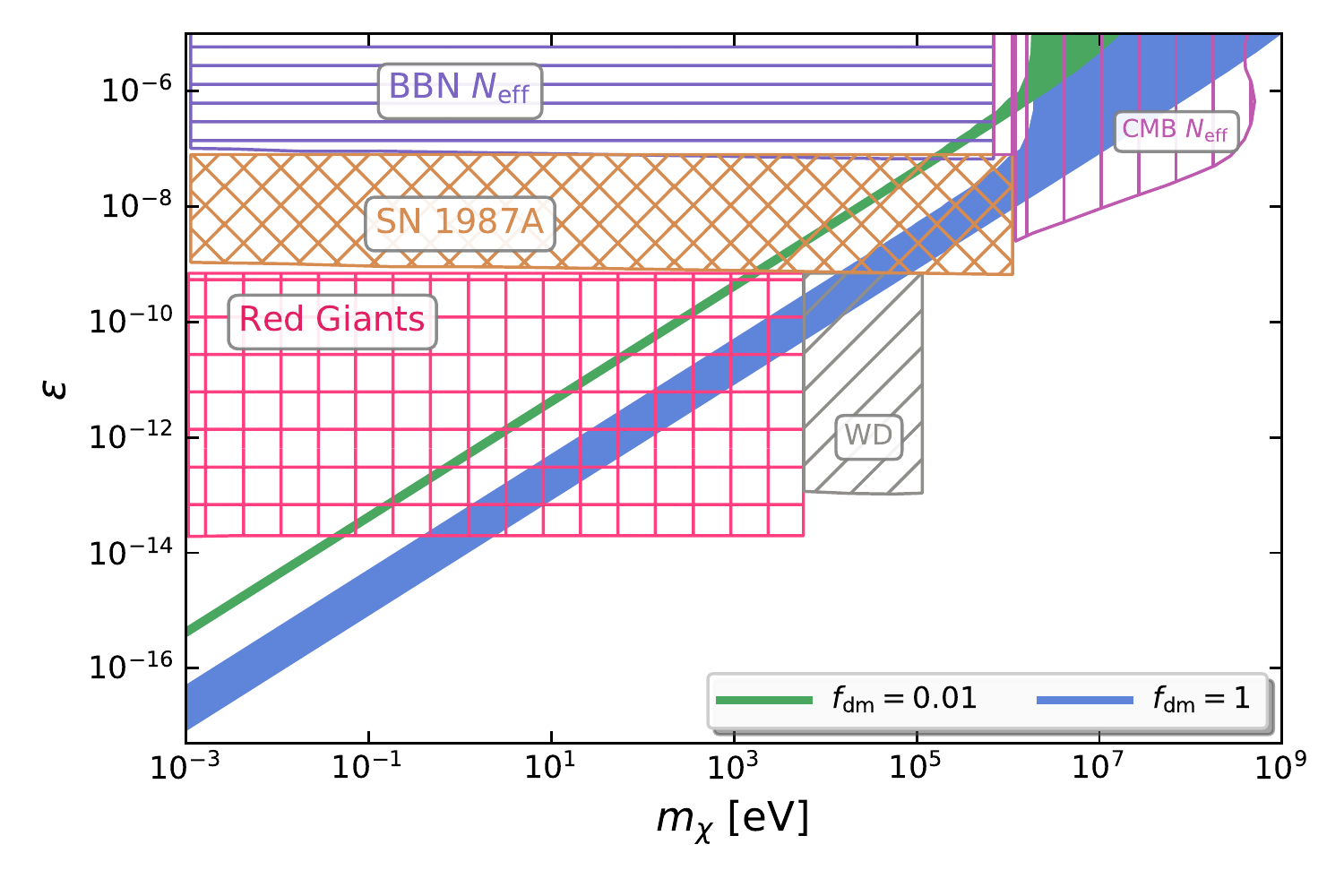}
 \caption{Constraints on the DM minicharge required
     to explain the EDGES signal.
     The blue (green) region represents our 99\% CL constraints
     for $f_{{\rm dm}}=1$ ($f_{{\rm dm}}=0.01$)
     that are consistent with our bound (\ref{eq:bounds})
     on the DM velocity dispersion.
     Bounds on MCPs from the early Universe and stellar physics
     are also shown (see text for references).
     The CMB-$N_\text{eff}$ bound applies only to the model
     that explicitly includes the hidden photon relativistic dof.
    }
\label{fig:minicharge}
\end{figure}

The strongest bounds on minicharge come from
\textit{the energy-loss argument}, alluding to the escape
of these particles from the core of stars
\citep{Raffelt1996book}.
Excitations of the dense electron-proton plasma
(also called plasmons) can decay into MCPs;
if the charge is low enough
\citep[$\epsilon \lesssim 10^{-8}$,][]{Davidson2000},
they propagate freely through the plasma 
and escape from the star \citep{Vinyoles2016}.
The dissipation of energy should modify
the usual stellar evolution,
thus limiting the plasmon decay-rate into MCPs,
and hence constraining $\epsilon$.
Combining studies of White Dwarfs (WD), Red Giants (RG),
the Super Nova 1987-A (SN87A), and the Sun (among others),
indicate that $\epsilon \lesssim 2\times 10^{-14}$
\citep[][see also Figure \ref{fig:minicharge}]{Davidson2000,Vinyoles2016,Chang2018SN87A}.

\begingroup
\setlength{\tabcolsep}{11pt} 
\renewcommand{\arraystretch}{1} 
\begin{deluxetable}{ccc}
\tablecaption{Bounds on the DM minicharge}
\tablehead{$f_\text{dm}$ & $m_\chi$ & $\epsilon$}
\startdata
  1    &  $10^{-3}-2$  eV & $ 8 \times 10^{-18} < \epsilon < 2\times10^{-14}$  \\
  0.1  &  $10^{-3}-0.4$ eV & $ 5 \times 10^{-17} < \epsilon < 2\times10^{-14}$ \\
  0.01 &  $10^{-3}-0.05$ eV & $ 4 \times 10^{-16} < \epsilon < 2\times10^{-14}$ 
\enddata
 \tablecomments{Bounds on $\epsilon$ are directly read from Fig. \ref{fig:minicharge}.}
\label{table:minicharge}
\end{deluxetable}	

In the DM mass range of this work (10$^{-3}$-10$^{9}$ eV),
the DM particle number density is always comparable or much larger than baryons.
Then, we should consider the cross-section due to
a baryon propagating in an MCP plasma
\citep{McDermott2011},
\begin{equation}
  \sigma_0 = \frac{2\pi\alpha^2\epsilon^2 \xi}{\mu_{\chi t}^2},
  \label{eq:sigma-epsilon}
\end{equation}
where $\alpha$ is the fine-structure constant and
$\mu_{\chi t}$ is the reduced mass between the DM and the baryon 
(proton or electron).
The Debye logarithm
$\xi = \log (9T_{\chi}^3/(4\pi\alpha^3 \epsilon^4 n_{\chi}))$,
which regulates the screening of the interaction by the plasma,
can be approximated in this case to
$\xi \approx  93 -4 \log(10^{14}\epsilon \; (\text{eV}/m_\chi))$.
Given that MCPs cannot interact with neutral atoms,
the energy transfer in equation (\ref{eq:heat-transfer}) 
gets suppressed by the fraction of free electrons.

We can now obtain minicharge bounds by inserting equation \eqref{eq:sigma-epsilon} into the heat transfer equation \eqref{eq:heat-transfer}, this is  depicted in Figure \ref{fig:minicharge}.
Notice that in order to simultaneously
explain the EDGES signal and avoid stellar bounds,
the DM mass needs to be towards the ultra-light regime.
From the non-excluded $m_\chi$–$\epsilon$ window,
some bounds on the DM minicharge
are listed in Table \ref{table:minicharge},
according to three values of $f_\text{dm}$.
Notice once again that a tighter bound on $\Delta\upsilon_0$
would result in a reduced allowed space for $\epsilon$.

Incidentally, notice that the scattering due to MCPs
is dominantly incoherent for our studied range of masses.
This is due to the smallness of $\epsilon$,
causing the MCP's mean free path $\ell_\chi=(\sigma(v)\,n_\chi)^{-1}$
to be extremely large compared to the energy-exchange length
$\lambda_{\chi b}=(\mu V_{\chi b})^{-1}$.
Despite the apparently high densities at lower masses
(\textit{e.g.} for $z=20$ and $m_\chi$=1 eV,
$n_\chi \sim 10^{10}$ cm$^{-3}$),
the smallness of $\epsilon$ makes the MCPs a rarefied plasma.
This might not be the case for ultra-light DM candidates.
A minicharge as small as $\epsilon \sim 10^{-14}$
will be enough to cause ($\ell_\chi < \lambda_{\chi b}$)
a scattering dominantly coherent
for $m_\chi \ll 10^{-6}$ eV.

\subsection{Electric dipole moment}
\label{sec:edm}
Following the same spirit of MCPs,
a type of neutral DM possessing an electric dipole moment
(EDM) $\mathcal{D}$ has been targeted for direct detection
\citep{Pospelov2000,Sigurdson2004,Sigurdson2006}%
\footnote{Magnetic dipole moments (MDMs) $\mathcal{M}$
have been experimentally targeted as well.
Here we do not consider MDMs because they produce
a velocity independent scattering ($n=0$) with baryons
\citep[see \textit{e.g.}][]{Sigurdson2006}.}.
These particles can only be Dirac fermions
in order to have a permanent dipole moment.
It is customary to report $\mathcal{D}$ in units of the Bohr magneton
$\mu_B=e\hbar/2m_e=1.93\times 10^{-11}\,e$ cm.

BBN sets an upper bound,
$\mathcal{D} \lesssim 5.2\times 10^{-12} \mu_B$,
in order to avoid late thermalization of particles below a few MeV's
\citep{Sigurdson2004}.
In the sub-GeV mass range, Collider Physics
is the most sensitive probe to DM-EDM 
through radiative corrections to the W boson mass,
which sets a mass independent upper limit at 
$\mathcal{D} \lesssim 1.6\times 10^{-5}\mu_B$, 
and from pertubative constraints from corrections to $Z$-pole observables,
requiring that $\mathcal{D} \lesssim 3.7\times 10^{-5}\mu_B$
\citep{Sigurdson2004}.
Stellar Physics constrain the neutrino magnetic dipole moment
from the energy-loss argument discussed above.
These constraints also apply to DM particles
coupling to photons through an EDM.
Accordingly, the most stringent astrophysical limits
correspond to the Sun, WD, RG, and SN87A, implying
$\mathcal{D} \lesssim 2\times 10^{-12} \mu_B$ 
\citep{Bertolami2014,Kadota2014,Arceo2015,Canas2016}.

\begin{figure}[b]
 \centering
 \includegraphics[width=\linewidth]{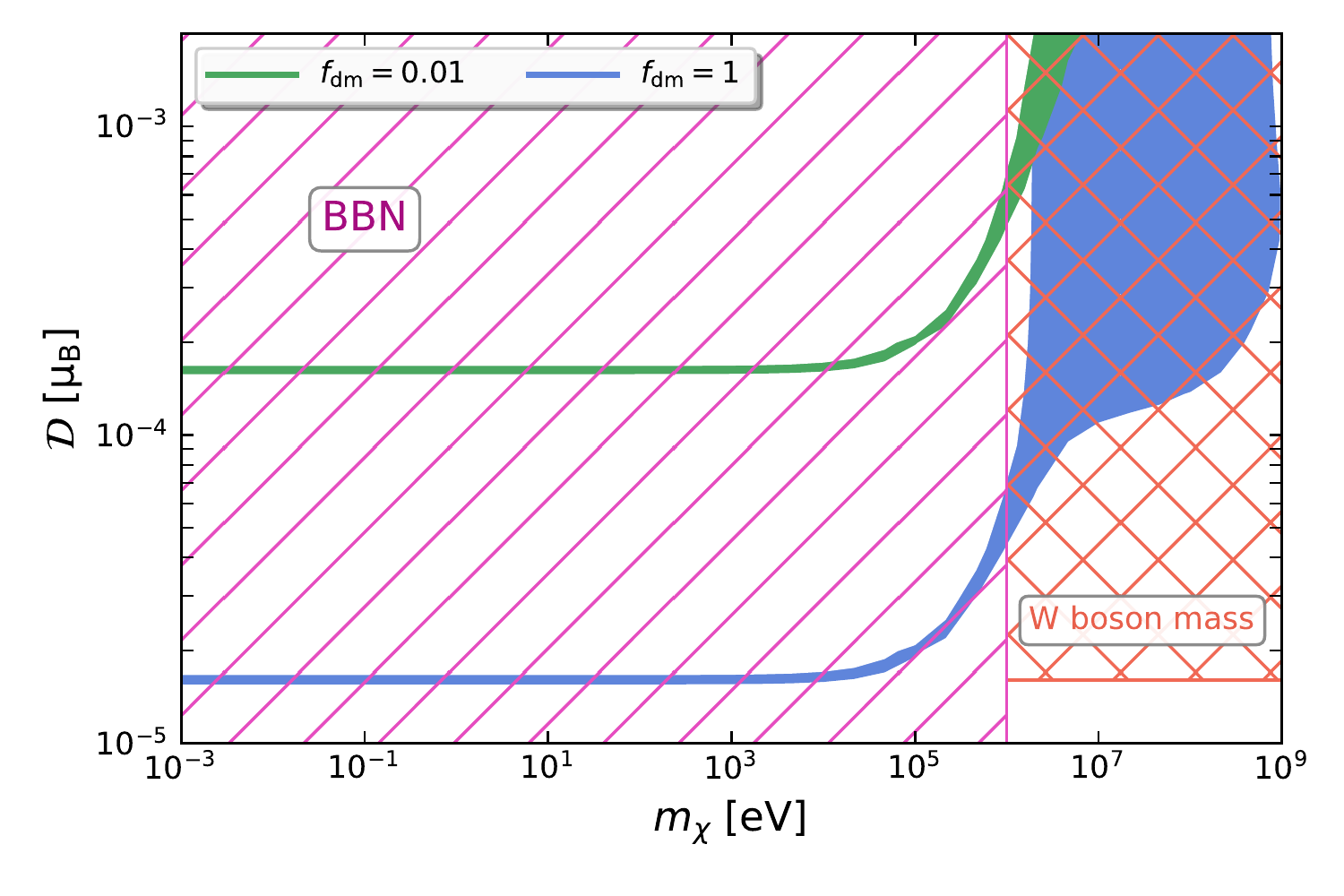}
 \caption{
     Constraints on the DM electric dipole moment
     required to explain the EDGES signal,
     along with the region already excluded by BBN and collider experiments.
     The blue (green) region represents the 99\% CL region consistent
     with our constraints on the DM velocity dispersion
     for $f_{{\rm dm}}=1$ ($f_{{\rm dm}}=0.01$).
     }
\label{fig:dipole}
\end{figure}

Our bounds on $\sigma_0$ computed with a $v^{-2}$ dependence
(as shown in Figure \ref{fig:sigma_m})
can be translated to $\mathcal{D}$, according to \citep{Sigurdson2004},
\begin{equation}
  \sigma_0 = 2 \alpha \mathcal{D}^2 .
\end{equation}
From the results depicted in Figure \ref{fig:dipole},
we can see that the EDM needed to explain the EDGES measurement
in the mass range $10^{-3}$–$10^{9}$ eV
is already discarded by the BBN constraints and
the measurements of the W boson mass in colliders.

As discussed above in the case of MCPs,
the scattering through an EDM is also dominated by incoherent scattering
in the mass range $10^{-3}$–$10^{9}$ eV.
For example, if $\mathcal{D}=10^{-4} \mu_B$,
the scattering would be coherent ($\ell_\chi < \lambda_{\chi b}$)
for $m_\chi \ll 10^{-6}$ eV.
For an EDM as small as $\mathcal{D}=10^{-12} \mu_B$,
the scale of coherent scattering is pushed down to $m_\chi \ll 10^{-14}$ eV.

\section{Conclusions}
\label{sec:close}

While the mass is quite an unknown aspect of DM,
its velocity dispersion is a physical property much less studied.
It is not uncommon to think that the DM relic velocity is either
necessarily zero (assuming Cold DM) or 
thermally suppressed by the particle mass (in Warm DM models),
like in equations (\ref{eq:velther1}) and (\ref{eq:velther2}).
Nevertheless, as we have reviewed, there might be plenty of
non-thermal mechanisms that would cause finite velocity dispersions,
to some degree disentangling velocity and mass.
Here, we have constrained a wide region of the
$m_\chi$–$\Delta\upsilon$ diagram (Fig. \ref{fig:mass_vel_disp})
by means of the linear regime of matter perturbations and
using current CMB and LSS data.
Our analysis provides useful upper limits to the DM
velocity dispersion, listed in Table \ref{tab:priors_results}
and summarized in equation (\ref{eq:bounds}).
In general, we have shown that DM particles can be considered as
Warm or Cold DM depending on their actual velocity dispersion,
irrespective of their mass.

As expected, active neutrinos and thermal axions
(Hot DM) are ruled out as the main source of DM.
Thermal majorons are found barely allowed by our constraints,
suggesting the need for further scrutiny with CMB and LSS data,
and possibly accounting for their non-thermal corrections.
Candidates for thermal DM are allowed 
above $\sim$100 eV's by our constraints,
while they are discarded for $m_\chi\lesssim$40 eV's,
even after considering non-thermal corrections.
Resonantly produced sterile neutrinos and
other non-thermal DM candidates are well inside our bounds.
Very light ($\ll$ 1 keV) DM particles are allowed by our constraints
as long as their velocity dispersion concurs with our bound in (\ref{eq:bounds}).
This motivates a deeper study of non-thermal production mechanisms
like those briefly discussed in \S\ref{sec:neutrinos}.
Heavy thermal candidates are well below our velocity constraints.

Our bound (\ref{eq:bounds}) on the DM velocity dispersion
is mainly limited by the maximum wavenumber
(0.2 h Mpc$^{-1}$) contained in the SDSS DR4 LRG data.
This motivates further studies using LSS data at smaller scales
like those from the Ly-$\alpha$ forest,
which can extend our analysis down to $k \lesssim 5 \text{ h Mpc}^{-1}$
and improve our constraints at least by an order of magnitude.

The DM velocity dispersion is a key ingredient of
the 21-cm dynamics at the epoch of the Cosmic Dawn.
If the anomaly in the absorption profile measured by EDGES
is to be explained by a baryon-DM interaction,
the $T_{21}$-signal by itself is not enough to constrain
both the DM relic velocity and the baryon-DM
scattering cross-section.
Hence, it is of great importance to investigate $\Delta\upsilon$
using independent techniques and sets of data.

In order to overcome the highest allowed velocities
and henceforth ensure an efficient baryon-cooling,
the values of $\sigma_0$ would need to be accordingly larger (as depicted in 
Figure \ref{fig:sigma_m}).
However, the largest $\sigma_0$-values
are in conflict with previous bounds
\citep{Slatyer2018early,Xu2018,Boddy2018critical}
obtained from CMB and LSS data.
This means that if DM particles are very heavy
($m_\chi \gg \text{1 MeV}$),
they ought to be initially really cold
in order to explain the EDGES observation.
If DM particles are very light ($m_\chi \ll \text{1 keV}$),
they do not seem to have tight restrictions 
on their initial velocities – other than (\ref{eq:bounds}) –
in order to explain both early and late cosmological data.
Yet, we speculate that such very-light DM scenarios
would in turn need a very early cooling mechanism
(like those discussed in \S\ref{sec:neutrinos})
in order to attain velocities much smaller than thermal candidates.

We conclude that the $\Delta\upsilon_0 > 0$ allowed values
found after our analysis can surely play a major role
in the phenomenology of baryon-cooling.
Once again, this motivates further studies with
Ly-$\alpha$ forest or other small-scale LSS data,
which could tighten the allowed parameter space for
the baryon-DM scattering cross-section and minicharge
(see figures \ref{fig:sigma_m} and \ref{fig:minicharge}).

Assuming that the $n=-4$ and $n=-2$ types of scattering
are due respectively to MCPs and EDMs,
our constraints on $\sigma_0$ translate to
novel bounding areas for $\epsilon$ and $\mathcal{D}$,
which are modified by our constraints on $\Delta\upsilon_0$.
This effect is interesting in general
for direct detection experiments at the low-energy end,
whose typical targets are WISPs
\citep{Jaeckel2010,Ringwald2012axionrev}.
In this direction of research,
a more complete and detailed sampling of the
($m_\chi$, $\Delta\upsilon_0$, $\sigma_0$)
parameter space would be needed.
In particular, the mass parameter space should be explored
considering that the interacting and non-interacting DM fractions
may be made of particles with different masses.
This characterization will involve explicitly collisional terms
in the baryon and DM Boltzmann equations,
and a Boltzmann hierarchy of differential equations.
Clearly, the former study would be very interesting
and represents one way to improve our analysis.

On a side note, we briefly mentioned the mass-scale of
incoherent/coherent scattering for MCPs and EDMs.
We lastly say that in the latter case,
the continuum nature of DM ought to be taken into account
in close similarity to scale-invariant scenarios \citep{Katz2016}.
Moreover, the associated multi-body interaction
might imply a scattering driven by higher-order multipoles,
possibly the anapole, quadrupole, or the DM polarizability
\citep{Pospelov2000,Ovanesyan2015}.

\acknowledgments
The authors thankfully acknowledge the computer resources provided by
the Laboratorio Nacional de Supercómputo del Sureste de México, 
CONACYT network of national laboratories.
This project was possible owing to partial support from
CONACYT research grants 237004, 490769,
F.C.~2016/1848, and FORDECYT-297324.
V.A-R. acknowledges partial support from the project CONACyT CB285721.
We also thank Tracy Slatyer, Cora Dvorkin,
R.E. Sanmiguel, and J.B. Muñoz for interesting discussions.
We especially want to thank an anonymous Referee for a critical review
that has led to a significant improvement of our paper.

\bibliography{references}

\end{document}